# Comparison of reverse current mechanisms in GaN Schottky diodes grown on sapphire versus ammonothermal GaN substrates

**B. Orfao[1,*], M. Abou Daher[1], H. Bouillaud[1], Y. Roelens[1], P. Prystawko[2], R. Kucharski[2], M. Bockowski[2] and M. Zaknoune [1]**

[1] IEMN, Institut d'Electronique de Microélectronique et de Nanotechnologie, Lille, France.
[2] Institute of High Pressure Physic, Polish Academy of Sciences, Sokolowska 29/37, Warsaw, 01-142, Poland.
E-mail: beatriz.orfao-e-vale-tabernero@univ-lille.fr

### Abstract

In this work, we analyse the reverse current mechanisms in GaN Schottky barrier diodes (SBDs) grown on sapphire and native GaN substrates. For the sapphire-substrate sample, two conduction mechanisms are identified: Poole–Frenkel emission (PFE) and trap-assisted tunneling (TAT), with corresponding trap energy levels of 0.9 eV and 0.3 eV, respectively. In contrast, only PFE is observed in the GaN-substrate sample, with a trap energy of 0.75 eV, suggesting that the presence of TAT is related to the higher dislocation density in structures grown on sapphire substrates. The leakage mechanisms and associated trap energies are extracted by comparing experimental current–voltage (I–V) characteristics with a model that includes thermionic emission and tunneling contributions for different temperatures, from 298K up to 443K.

## 1. Introduction

Wide bandgap (WBG) semiconductors present material properties that significantly improve the conventional technologies. Between them, we highlight gallium nitride (GaN) which presents a wide bandgap, high breakdown field, electron velocity and thermal conductivity [1]. These characteristics make GaN a promising candidate for high power, high frequency and high temperature applications.

Schottky diodes are widely used in frequency multipliers and mixers because of their strong non-linearity and in parallel, their high switching speed enables their use in fast-response applications [2,3]. While GaAs remains the most widely used technology for Schottky diodes, its limited thermal stability, relatively low breakdown voltage and modest power handling capability make GaN a strong candidate for more demanding applications. GaN Schottky barrier diodes (SBDs) are particularly well suited for high-power, high-temperature and space electronics, and additionally benefit from the intrinsic radiation hardness of GaN.

Reverse leakage current in GaN SBDs is comonly higher than that expected from thermionic and tunnel currents [4]. This excess current is generally associated with the presence of defects. The reverse current is often attributed to defect-assisted mechanisms such as Poole Frenkel emission (PFE), trap assisted tunneling (TAT), and variable range hopping (VRH) [5-6]. As a consequence, device reliability remains a key challenge that must be thoroughly investigated, as it is a critical requirement for applications such as space, medical, and security systems, where stable and long term operation is essential.

The lack of native GaN substrates has promoted the use of foreign substrates such as Si, SiC or sapphire. Although GaN layers grown on sapphire and SiC substrates exhibit a high dislocation density, typically ranging from $10^7$ to $10^9$ cm$^{-2}$, they are widely used due to their low cost, large wafer availability, and technological maturity [7]. However, the growth of GaN on native GaN substrates, which exhibit a much lower dislocation density of approximately $10^4$ to $10^5$ $cm^{-2}$, has significantly evolved in recent years [8].

The analysis of non-ideal current mechanisms has already been analysed in GaN SBDs by using temperature-dependent electrical characterization such as current-voltage (I-V) and Capacitance-voltage (C-V) measurements [9-11]. These studies have demonstrated the presence of PFE, TAT and VRH mechanisms. However, to the best of our knowledge, no comparison between substrates has been reported in the literature.

The aim of this work is to analyse the non-ideal current mechanisms present in GaN SBDs and to investigate their differences as a function of the substrate. The article is organized as follows. In section 2, we include the structures of the measured diodes. In section 3, the results are discussed to identify the non-ideal current mechanisms. Finally, in section 4 the main conclusions of our work are summarized.

## 2. Epitaxy and fabrication process

The diodes under analysis were fabricated on two similar epitaxial structures. The first one consists of a 600 nm thick epilayer (with a doping of $5\times10^{16}$ $cm^{-3}$), a 500 nm highly Si-doped GaN ($10^{19}$ $cm^{-3}$) and a 1.2 µm semi-insulating GaN buffer on a sapphire substrate. The second one has, from top to bottom, a 620 nm drift layer doped at $8\times10^{16}$ $cm^{-3}$, then a 425 nm GaN layer (with doping of $10^{19}$ $cm^{-3}$), and a 426 nm insulating GaN buffer layer on a GaN ammonothermal substrate.

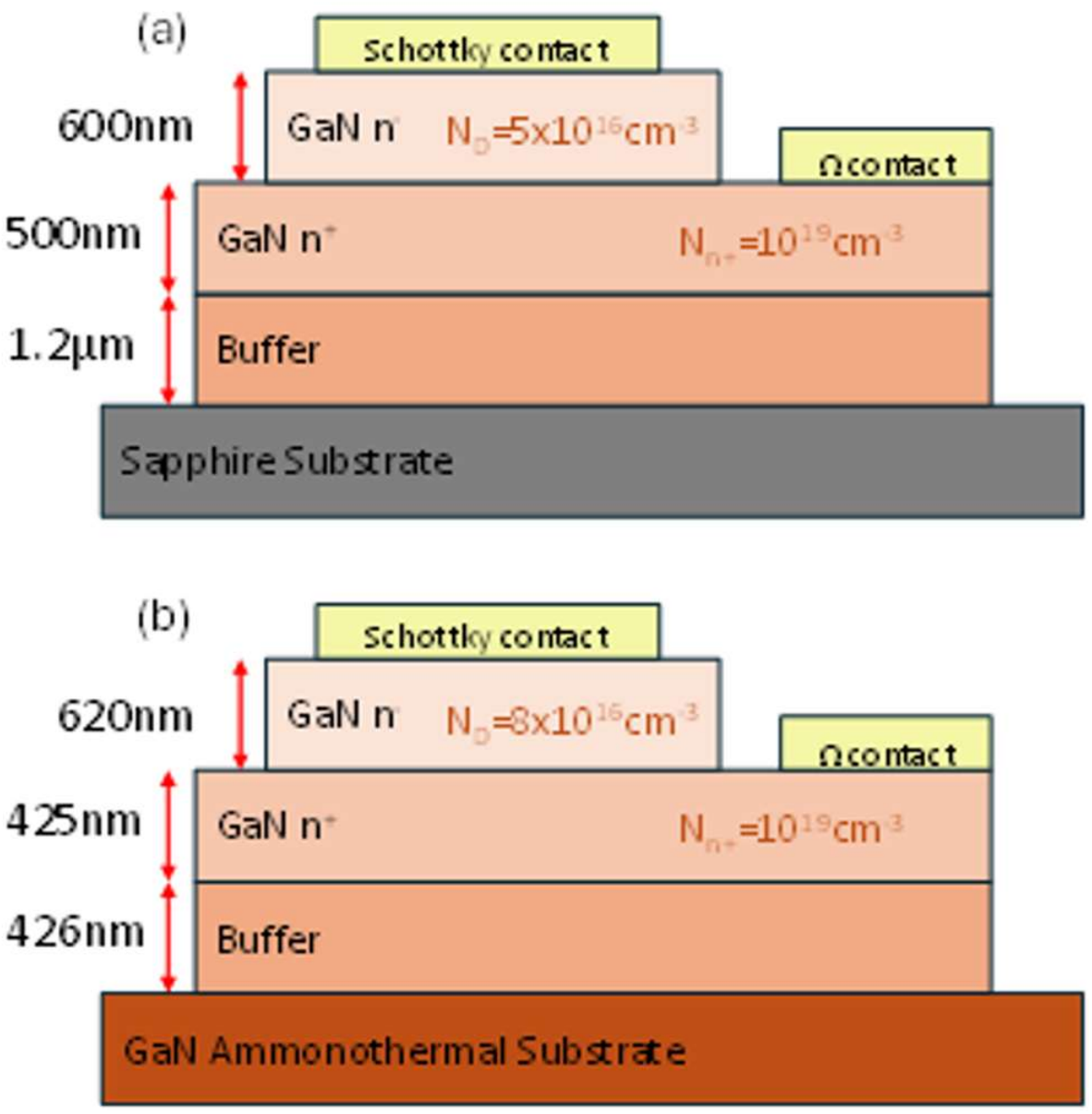

**Figure 1.** Scheme of the fabricated GaN SBDs on (a) sapphire and (b) GaN ammonothermal substrates.

The Schottky contact is placed on top of the epilayer, and the metal stack used is Pt/Au, while the ohmic contact, formed by Ti/Al/Ni/Au, is on the highly doped GaN layer. A $SiO_2$ hard mask was employed to define the mesa, followed by etching of the drift layer using a $Cl_2$/Ar inductively coupled plasma (ICP). For the sapphire-substrate sample, both ohmic and Schottky contacts were simultaneously annealed at 550 °C for 15 min in $N_2$ ambient. For the ammonothermal-substrate sample, the ohmic contact annealing was performed at 850 °C for 30 s in $N_2$ ambient and the Schottky contact at 400 °C for 20 min. A SEM image of the final state of the fabricated diodes is shown in figure 2.

The measurements presented in this work were carried out in circular diodes of 140 µm of diameter.

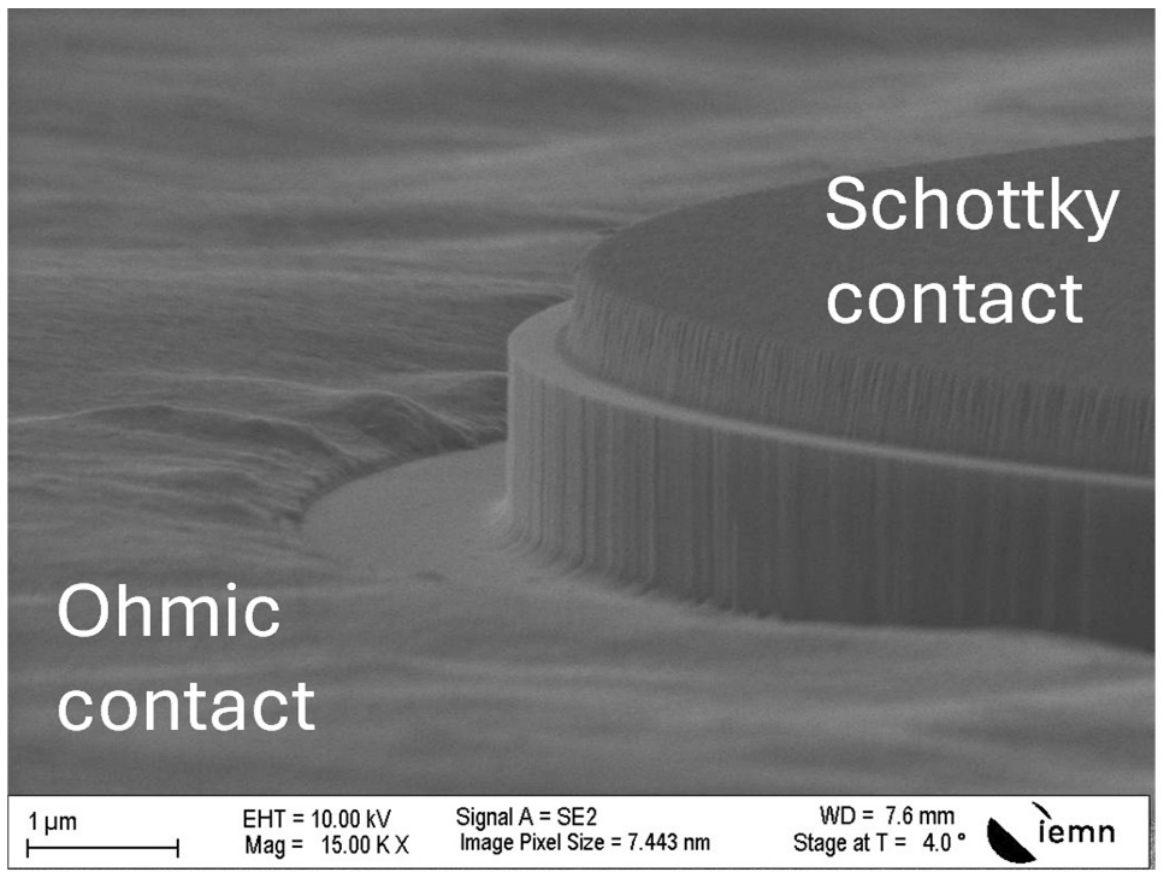

**Figure 2.** SEM image of the fabricated GaN Schottky diodes.

## 3. Results

C-V measurements have been performed at room temperature to extract the doping level of the drift layer ($N_D$) by using the following equation:

$$(C)^{-2} = \frac{2(V_B - V)}{S^2 q N_D \varepsilon_{sc}},$$

where S is the Schottky contact surface, q is the electron charge, $\varepsilon_{SC}$ is the permittivity of the semiconductor, $V_B$ is the built-in voltage. We obtained $N_D$ of $1.9\times10^{16}$ $cm^{-3}$ and $1.1\times10^{17}$ $cm^{-3}$ for the sapphire-substrate and GaN-substrate samples, respectively.

I-V measurements have been carried out at different temperatures, from 25ºC up to 170ºC by using a homemade experimental setup. The experimental setup consists of a power supply used to apply a voltage to a ceramic heater mounted beneath a metallic chuck, which serves as a heated sample holder. The temperature of the holder is monitored using an Agilent 34461A digital multimeter connected to a temperature sensor placed directly on the metallic chuck. Electrical probing is performed using tungsten probe tips, chosen for their thermal and mechanical stability under high temperature conditions.

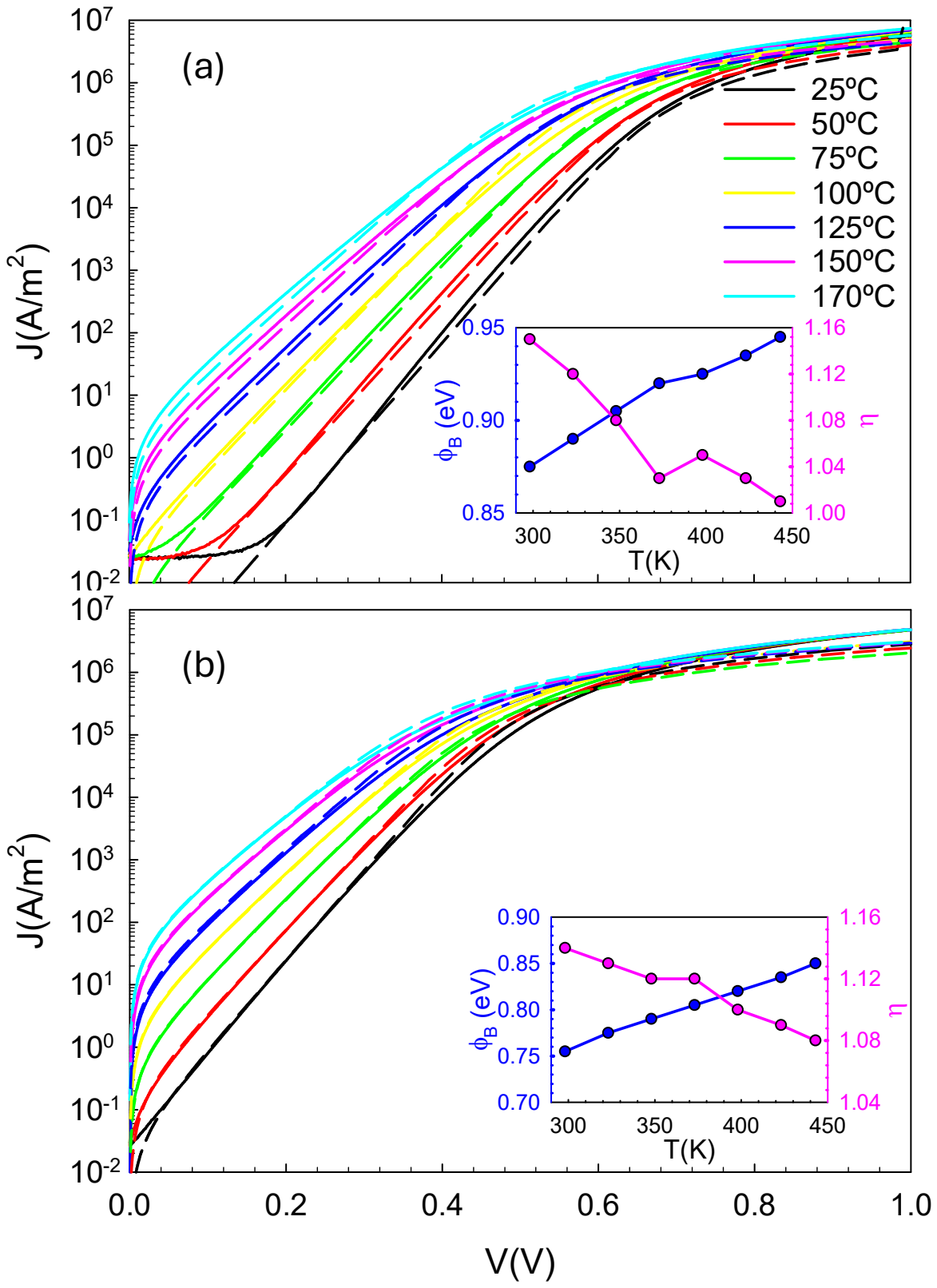

**Figure 3.** Forward I-V characteristics for the measurements (solid lines) and the model (dashed lines) for (a) sapphire-substrate and (b) GaN-substrate. The insets include $\phi_B$ and $\eta$ extracted from the fitting of the experimental curves and the model.

Figure 3 shows the forward I-V characteristics for (a) sapphire-substrate and (b) GaN-substrate, jointly with the results of the model developed in Ref. [12]. The model considers only the thermionic emission in forward bias while in reverse bias we take into account the thermionic and tunneling currents. The derivation of the expressions used to calculate these currents is detailed in Ref. [12].

The values of $\phi_B$ and $\eta$ have been extracted by fitting the experimental curves and the ideal current obtained from the model. We observe in the insets of figure 3 that, for both samples, the ideality factor decreases while the barrier height increases with increasing temperature. This behavior is typically attributed to Schottky barrier inhomogeneities [13,14].

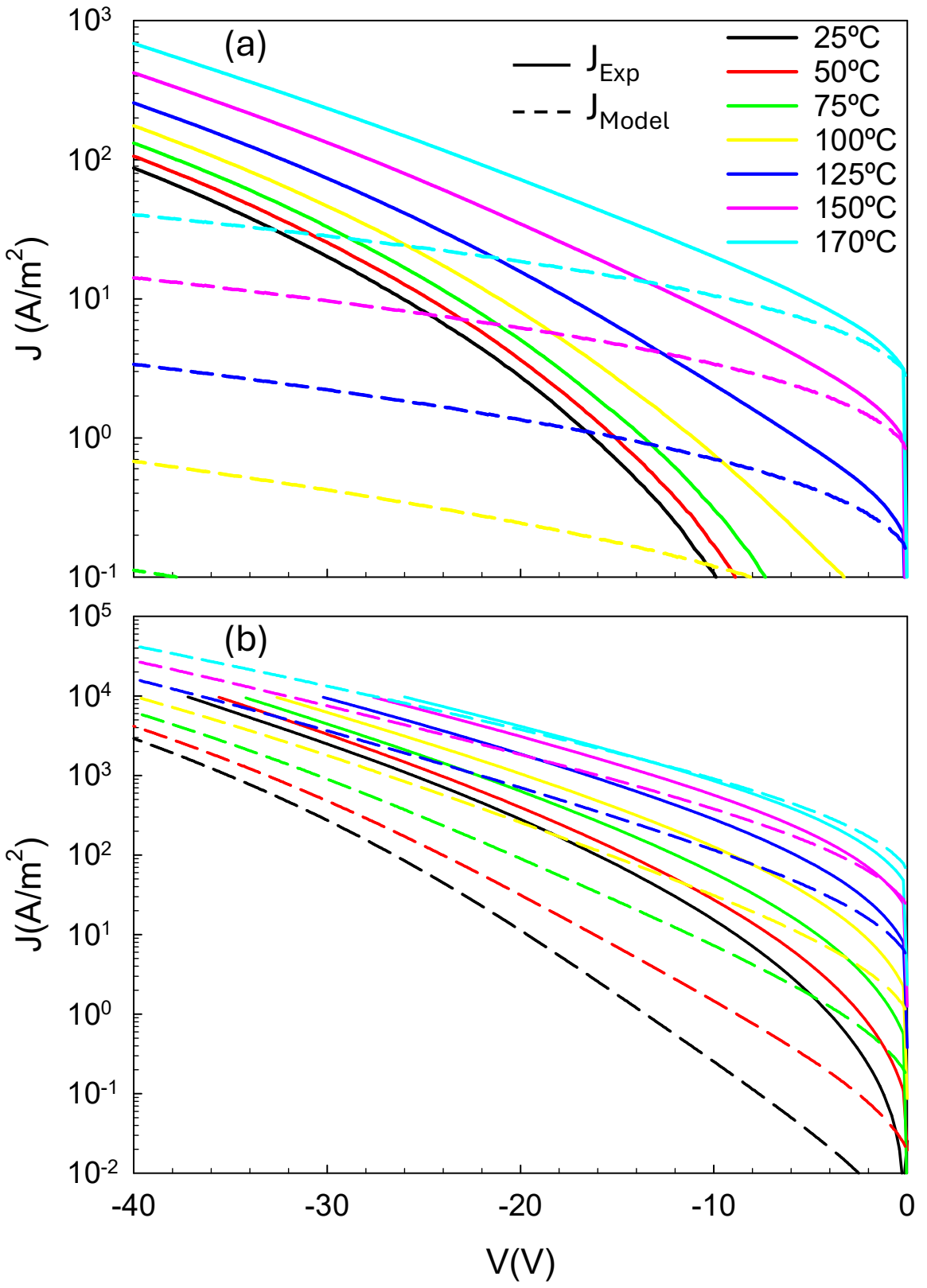

**Figure 4.** Reverse I-V characteristics for the measurements (solid lines) and the model (dashed lines) for (a) Sapphire-substrate and (b) GaN-substrate.

Figure 4 represents the reverse I–V characteristics obtained from both experimental measurements and the model. For the diode fabricated on a sapphire substrate, a significant difference between the measured current and the ideal current is observed over the entire temperature range. In contrast, for the GaN-substrate sample, the experimental curves approach the ideal current at the highest temperatures.

As we have previously mentioned, the reverse ideal current is estimated by the thermionic and tunneling contributions. The deviation from the ideal current indicates the presence of non-ideal current mechanisms. This deviation is more pronounced for the sapphire-substrate sample, and the difference between the measured current and the ideal current determines the excess current associated with non-ideal mechanisms.

The excess of current is used to identify the additional mechanisms present in each structure. By representing

Ln(J/E) as a function of √E a linear behavior is obtained for both samples indicating the presence of PFE. The current density associated with PFE can be expressed as [6]:

$$J_{PFE} \propto E exp\left(\frac{-q\left(\phi_t - \sqrt{qE/\pi\varepsilon_{sc}}\right)}{k_B T}\right),$$

where $E$ is the electric field, $k_B$ is the Boltzmann constant, $T$ is the temperature, $\phi_t$ is the trap energy level and $\varepsilon_{sc}$ is the high frequency dielectric constant ($5.35\varepsilon_0$ for GaN) [15,16]. Consequently, the linear dependence observed in Figure 5 confirms the contribution of PFE. The slope of the linear fit is expected to be equal to $q\sqrt{q/\pi\varepsilon_{sc}}/k_B T$. The slope obtained for the sapphire-substrate sample is higher than the expected value, indicating the contribution of additional conduction mechanisms, whereas for the GaN-substrate sample it is close to the theoretical value. Therefore, for the GaN-substrate sample, Poole–Frenkel emission is the dominant conduction mechanism, with an extracted trap energy level of approximately 0.75 eV.

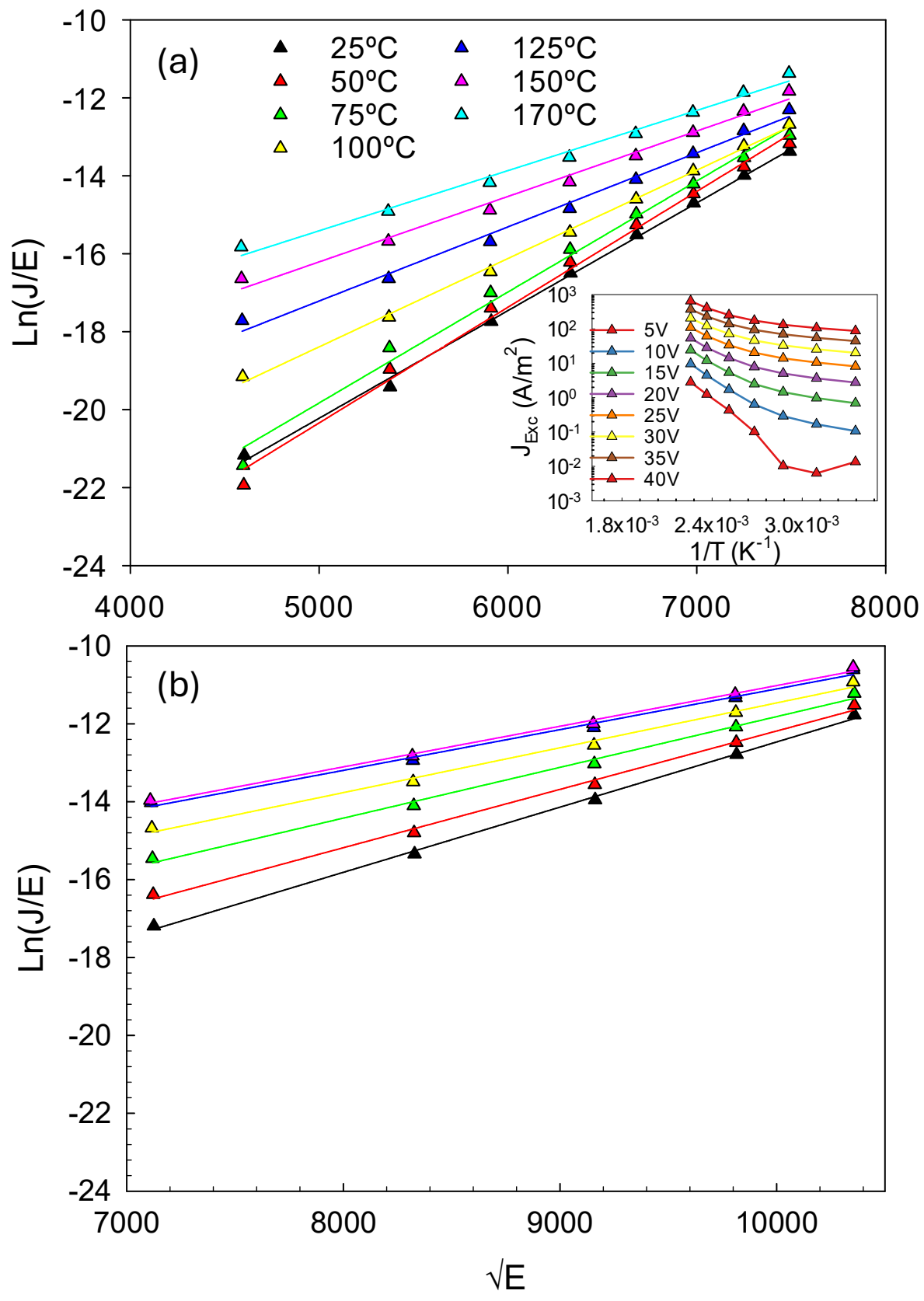

**Figure 5.** Representation of Ln(J/E) versus √E, which is the typical dependence of PFE mechanisms, including the fitting for (a) sapphire-substrate and (b) GaN-substrate diodes. The inset represents the excess current as a function of the inverse of temperature for the diode fabricated on sapphire substrate.

We next focus on identifying and analyzing the conduction mechanisms responsible for the excess current observed in the sapphire-substrate sample. The inset of Figure 5 shows the excess current as a function of the inverse temperature for the sapphire-substrate sample. From this representation, we observe a linear dependence at higher temperatures, while at lower temperatures the current tends toward a constant value, independent of temperature. This suggests that at lower temperatures, TAT may dominate, whereas at higher temperatures a combination of PFE and TAT likely contributes to the larger slope observed in Figure 5(a).

The current density associated with TAT is given by the expression [6]:

$$J_{TAT} \propto exp\left(\frac{-4\sqrt{2qm^*}(\phi_t)^{3/2}}{3\hbar E}\right),$$

where m* is the effective mass of the electrons ($0.22m_0$ for GaN) [17] and $\hbar$ is the Planck constant.

At lower temperatures and lower electric fields, where the current becomes nearly independent of the temperature, the trap energy level responsible for TAT can be determined by representing Ln(J) as a function of 1/E, see Figure 6. The trap energy level obtained is 0.29 eV. By determining the TAT current, we can estimate the contribution of this mechanism to the total excess current, and subsequently obtain the contribution associated with PFE using $J_{PFE} = J_{Exc} - J_{TAT}$. Therefore, by plotting Ln($J_{PFE}$/E) as a function of √E at high temperatures and high electric fields (see Figure 6), we observe that the slope approaches the expected theoretical value, and the corresponding trap energy level can be calculated to be approximately 0.9 eV.

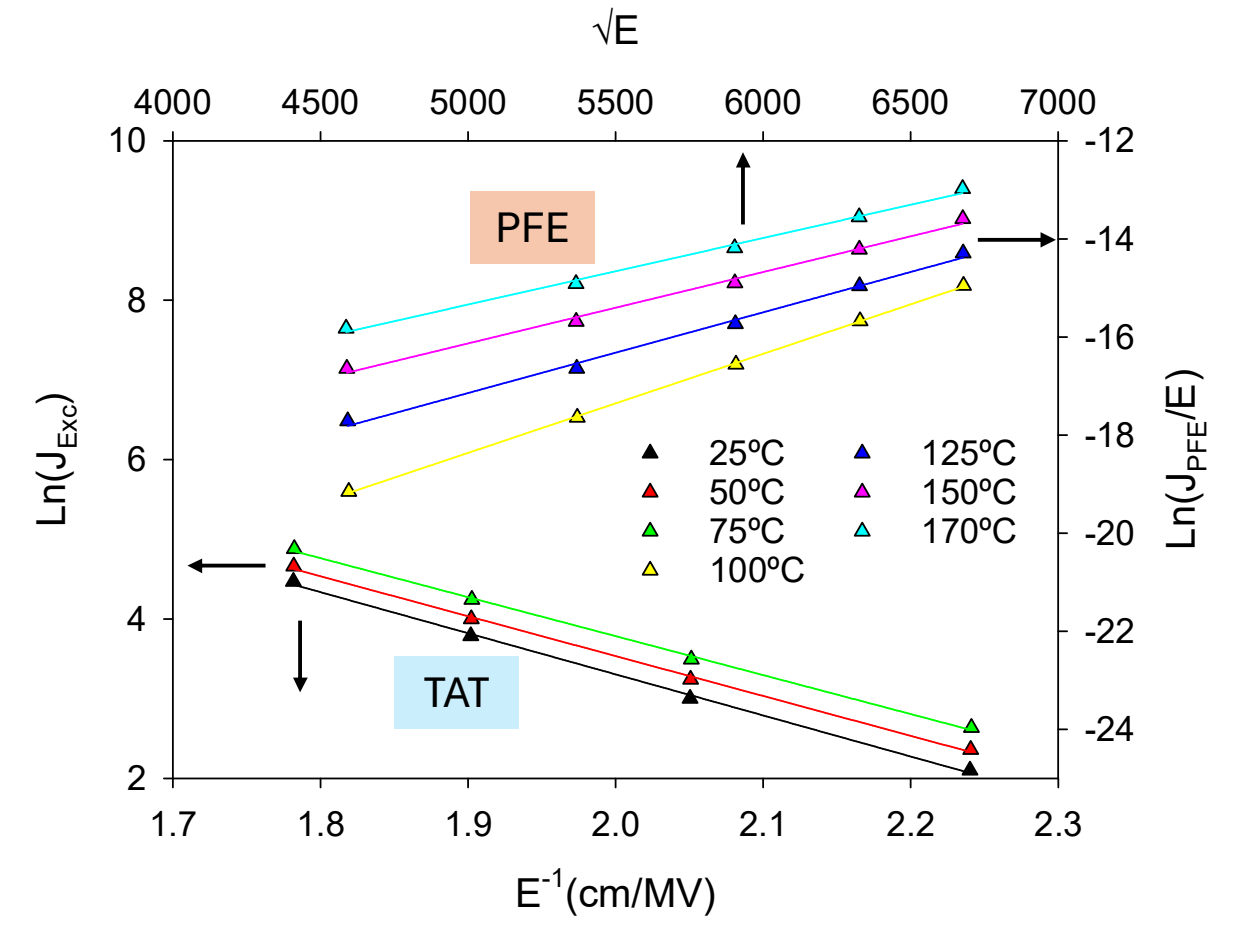

**Figure 6.** Representation of the contributions of Poole–Frenkel emission (PFE) and trap-assisted tunneling (TAT) to the excess current in the sapphire-substrate sample.

The fact that only PFE is observed in the GaN-substrate sample may be attributed to its low dislocation density, typically $10^4$ to $10^5$ $cm^{-2}$. In contrast, for the sapphire substrate, the dislocation density is much higher ($10^7$ to $10^9$ $cm^{-2}$) and explains the presence of both PFE and TAT. Regarding the PFE mechanism, it is reasonable to observe it at high temperatures and high electric fields, since the traps involved are deep (with energies of 0.75 eV and 0.9 eV) and electrons require sufficient thermal energy to escape from these traps. These deep traps are generally associated with nitrogen vacancies [18,19].

## 4. Conclusions

For the sapphire-substrate sample, both Poole–Frenkel emission (PFE) and trap-assisted tunneling (TAT) are observed. At high temperatures and high electric fields, PFE dominates with a trap energy of 0.9 eV, whereas at lower temperatures and lower fields, TAT is the dominant mechanism with 0.29 eV. In contrast, for the GaN-substrate sample, only PFE is observed, with a trap energy of approximately 0.75 eV. This difference between the samples may be attributed to the significantly lower dislocation density and reduced defect concentration in the GaN substrate, which suppresses the contribution of tunneling-assisted mechanisms and results in a single dominant PFE process.

## Acknowledgements

This work is supported by CPER Wavetech and Imitech. The nano-fabrication is supported by the French network RENATECH, and the Equipex+ Nanofutur program. The substrate and epitaxy part of this work has been also supported by the Wide Bandgap (WBG) Pilot line, which is funded jointly by the Chips Joint Undertaking, through the European Union's Digital Europe programme and Horizon Europe programme, as well as by the participating states Italy, Sweden, Poland, Finland, Austria, France and Germany, under Grant Agreement n. 101183211.

## References

[1] Sun Y, et al. 2019 *Electronics* **8** 5.
[2] Zhang B, Ji D, Fang D, Liang S, Fan Y and Chen X 2019 *IEEE Electron Device Lett.* **40** 780–783.
[3] Liang S, Song X, Zhang L, Lv Y, Wang Y, Wei B, Guo Y, Gu G, Wang B, Cai S, and Feng Z 2020 *IEEE Electron Device Lett.* **41** 669–672.
[4] Hashizume T, Kotani J and Hasegawa H 2004 *Appl. Phys. Lett.* **84** 4884.
[5] Miller E J, Yu E T, Waltereit P and Speck J S 2004 *Appl. Phys. Lett.* **84** 535.
[6] Fu K, Fu H, Huang X, Yang T H, Cheng C Y, Peri P R, Chen H, Montes J, Yang C, Zhou J, Deng X, Qi X, Smith D J, Goodnick S M and Zhao Y 2020 *IEEE J. Electron Devices Soc.* **8** 74–83.
[7] Sakai S, Wang T, Morishima Yand Naoi Y 2000 *Journal of crystal growth* **221** 334-337.
[8] Yoshida T an Shibata M 2020 *Japanese Journal of Applied Physics* **59** 071007.
[9] Pipinys P and Lapeika V 2006 *J. Appl. Phys.* **99** 093709.
[10] Chen J, Liu Z, Wang H, Song X, Bian Z, Duan X, Zhao S, Ning J, Zhang J and Hao Y 2021 *Appl. Phys. Express* **14** 104002.
[11] Orfao B, Abou Daher M, Peña R A, Vasallo B G, Pérez S, Íñiguez-De-La-Torre I, Paz-Martinez G, Mateos J, Roelens Y, Zaknoune M and González T 2024 *Journal of Applied Physics* **135** 1.
[12] Orfao B, Di Gioia G, Vasallo B G, Pérez S, Mateos J, Roelens Y, Frayssinet E, Cordier Y, Zaknoune M and González T 2022 *Journal of Applied Physics* **132** 4.
[13] Ejderha K, Duman S, Nuhoglu C, Urhan F and Turut A 2014 *J. Appl. Phys.* **116** 234503.
[14] Gülnahar M 2014 *Superlattices Microstruct.* **76** 394–412.
[15] Simmons J G 1967 *Physical Review* **155** 657.
[16] Madelung O 2004 *Semiconductors: Data Handbook* Springer, Berlin.
[17] García S, Pérez S, Íñiguez-De-La-Torre I, Mateos J and González T 2014 *J. Appl. Phys.* **115** 044510.
[18] Maurya V, Buckley J, Alquier D, Irekti M R, Haas H, Charles M, Jaud M A and Sousa V 2023 *Energies* **16** 5447.
[19] Peta K R, Park B G, Lee S T, Kim M D, Oh J E, Kim T G, Reddy V R 2013 *Thin solid films* **534** 603-608.